\documentclass[preprint,12pt]{elsarticle}

\usepackage{amssymb}
\usepackage{amsmath}
\usepackage{bm}

\usepackage{booktabs}
\usepackage{multirow}
\usepackage[table]{xcolor}
\usepackage{array}
\usepackage{tabularx}
\usepackage{siunitx}
\usepackage{threeparttable}

\usepackage{graphicx}

\usepackage{algorithm}
\usepackage{algpseudocode}

\usepackage{caption}

\begin{document}



\begin{frontmatter}



\title{Deep learning framework for predicting stochastic take-off and die-out of early spreading}

\author[aff1]{Wenchao He} 
\author[aff1,aff3]{Tao Jia\footnote{Corresponding author: tjia@swu.edu.cn}}

\affiliation[aff1]{organization={College of Computer and Information Science},
            addressline={Southwest University}, 
            city={Beibei},
            postcode={400715}, 
            state={Chongqing},
            country={People's Republic of China}}

\affiliation[aff3]{organization={College of Computer and Information Science},
            addressline={Chongqing Normal University}, 
            city={},
            postcode={401331}, 
            state={Chongqing},
            country={People's Republic of China}}
            
\begin{abstract}

Large-scale outbreaks of epidemics, misinformation, or other harmful contagions pose significant threats to human society, yet the fundamental question of whether an emerging outbreak will escalate into a major epidemic or naturally die out remains largely unaddressed. This problem is challenging, partially due to inadequate data during the early stages of outbreaks and also because established models focus on average behaviors of large epidemics rather than the stochastic nature of small transmission chains.
Here, we introduce the first systematic framework to address the stochastic take-off prediction problem-forecasting whether initial transmission events will stochastically amplify into major outbreaks or fade into extinction during early stages, when intervention strategies can still be effectively implemented.
Leveraging extensive data generated from stochastic spreading models, we developed a deep learning-based framework capable of predicting the stochastic take-off and die-out of early-stage spreading events in real-time. We validate the effectiveness of our machine learning framework by examining spreading scenarios with varying levels of infectivity on both Erdős-Rényi (ER) and Barabási-Albert (BA) networks. We find that the proposed deep learning-based method achieves accurate predictions for stochastic spreading well in advance of potential outbreaks and demonstrates significant robustness across different infectivity scenarios and various network structures.
To address the challenge of sparse data during early outbreak stages, we further propose a pretrain-finetune framework that leverages diverse simulation data for pretraining and adapts to specific scenarios through targeted fine-tuning, significantly enhancing cross-domain generalization capability.
The pretrain-finetune framework consistently outperforms baseline models, achieving superior performance even when trained on limited scenario-specific data. 
To our knowledge, this work presents the first framework for predicting stochastic take-off versus die-out.
This framework provides valuable insights for epidemic preparedness and public health decision-making, enabling more informed early intervention strategies.

\end{abstract}


\end{frontmatter}




\section{Introduction}

The accurate prediction of information spreading, innovation diffusion, or disease transmission within populations, whether in digital or physical environments, remains a fundamental challenge across sociology, epidemiology, and informatics \cite{jackson2007diffusion, ugander2012structural, cheng2014can, rosenthal2015revealing, scarpino2019predictability, estrada2020covid}. These diverse spreading processes, ranging from technological adoption to misinformation spread and viral infections, share an inherent unpredictability \cite{salganik2006experimental,bertozzi2020challenges,rosenkrantz2022fundamental,castro2020turning,wilke2020predicting}.
This unpredictability stems not only from incomplete knowledge (epistemic uncertainty) but also from aleatoric uncertainty – the intrinsic randomness inherent in transmission events due to individual-level variability, which cannot be reduced even with perfect information \cite{der2009aleatory,penn2023intrinsic}.  Each transmission event essentially becomes a probabilistic outcome governed by the complex interplay of many random factors, making early-stage predictions particularly sensitive to stochastic fluctuations.

This predictive challenge is particularly evident in the context of contagious diseases. One of the extremely challenging problems is determining whether a number of early detected cases suggest a single tree of a few local transmissions that will naturally die out, or portend a large-scale outbreak with catastrophic consequences. Despite decades of epidemic modeling, the research community was ill-equipped to solve this problem. The issue is challenging partially due to the lack of efficient data in the early times of the epidemic outbreak, and also because the well-established epidemic models mainly operate based on deterministic mechanisms,  which are designed to describe the average behavior of large epidemics, rather than the stochastic and heterogeneity nature of early transmission chains. 

The deterministic models commonly assume that the populations in the various compartments are homogenous, in the sense that all individuals behave similarly, and are well-mixed \cite{kermack1927contribution,pastor2015epidemic}. These models can be useful for understanding the overall dynamics of an epidemic and provide a valid approximation for some applications \cite{pastor2015epidemic, cai2022modeling}. 
However, at the beginning of the spreading process, the number of infectious individuals is typically small, and the transmission is dominated by stochastic fluctuation rather than deterministic trends \cite{czuppon2021stochastic, britton2019stochastic, thompson2016detecting}. 
This uncertainty manifests through variations in individual health conditions, disease transmissibility, and contact patterns within the population \cite{castro2020turning}.
The heterogeneous nature of disease transmission is well manifested in superspreading phenomena \cite{lloyd2005superspreading}, where a minority of infected individuals account for a disproportionate majority of transmission events. For instance, empirical analyses of COVID-19 transmission clusters reveal that 80\% of secondary infections stemmed from merely 10-20\% of index cases, with most infected individuals generating few subsequent transmissions \cite{lemieux2021phylogenetic, sun2021transmission, sneppen2021overdispersion}. 
These observations suggest that assumptions of homogeneous mixing during early epidemic stages may be unrealistic, rendering deterministic models insufficient. Accurate modeling and prediction of early-stage outbreak dynamics necessitate frameworks that incorporate both network structural properties and stochastic processes \cite{dangerfield2009integrating,lloyd2005superspreading,andersson2012stochastic,gillespie1977exact,allen2015stochastic}.

In this paper, we examine the various spreading patterns through stochastic epidemic modeling that incorporates both heterogeneity and stochasticity. More importantly, leveraging the extensive simulation data generated from these stochastic models, we developed a deep-learning-based framework capable of predicting the stochastic take-off and die-out of an early spreading.  We found that our proposed Outbreak-GWN model, which predicts outbreaks by learning the structural and temporal information simultaneously, demonstrates remarkable performance in predicting outbreaks in their early stages. 
We validate the effectiveness of our machine-learning framework by considering the spreading in scenarios with varying levels of infectivity on both the Erdős–Rényi (ER) network and the Barabási–Albert (BA) network. We find that the proposed deep-learning based method is capable of making accurate predictions for stochastic spreadings well ahead of their outbreaks, and demonstrates significant robustness and generalizability across different infectivity scenarios and various network structures. 
To address sparse or unknown parameters in early outbreak stages, we propose a pretrain-finetune framework that leverages diverse simulation data for pretraining and adapts to specific scenarios through targeted fine-tuning, significantly enhancing cross-domain generalization. This pretrain-finetune method consistently outperforms other baselines, achieving superior performance even with limited scenario-specific data.

To our knowledge, this work represents the first systematic framework for early-stage stochastic outbreak prediction that addresses the fundamental challenge of distinguishing between stochastic die-out and take-off dynamics. Our findings advance the understanding of stochastic epidemic patterns and present a comprehensive framework for predicting stochastic outbreaks. 
\label{sec:introduction}

\section{Take-off and die-out in the spreading}

To model the spreading of contagions, we adopt the well-known susceptible-infected-recovered (SIR) model \cite{kermack1927contribution,colizza2006role,pastor2015epidemic,geng2021kernel}, where individuals in a population are categorized according to their infection status: susceptibles (S), infectious (I), or recovered (R).  The deterministic spreading models are usually formulated by a set of ordinary differential equations (ODEs). The standard deterministic SIR model is given by: 
\begin{equation}
\begin{aligned}
&\frac{dS}{dt} =-\frac\beta NSI \\
&\frac{dI}{dt} =\frac\beta NSI-\mu I \\
&\frac{dR}{dt} =\mu I,
\end{aligned}
\end{equation}
where \textit{N} represents the total population size, $\beta$ is the transmission rate, and $\mu$ is the recovery rate.   The deterministic SIR model is widely used to describe the average behavior of large epidemics and provide an approximation for many applications \cite{pastor2015epidemic, cai2022modeling}. However, the pre-determined nature of deterministic models implies that they would fail to account for the inherently stochastic nature of disease spread, which is particularly important in the early stages of a spreading process, as the infection numbers will be few and so random variations alone can cause spreading to die out or take off \cite{lloyd2005superspreading,  thompson2016detecting,castro2020turning, penn2023intrinsic}.  
Fig. 1(a) depicts the trajectories of 200 stochastic simulation samples generated by the stochastic SIR simulations and the dynamics of the corresponding deterministic SIR model. As shown in Fig. 1(a), in contrast to the deterministic SIR model, which eventually converges to a globally stable equilibrium point, the stochastic simulations show that there are generally two kinds of trajectories. In one scenario, a few individuals are infected and then the transmission is eventually ended; in the other scenario, a fairly large fraction of individuals are infected. And there are almost no trajectories between these two outcomes. Fig. 1(b) presents the distributions of their final recovered numbers from a larger number (100,000) of stochastic simulations. The distribution exhibits a bimodal behavior in the spreading regime.  The local phase, which approaches 0, is referred to as stochastic die-out, while those corresponding to the other peak are referred to as stochastic take-off or outbreak. 
\begin{figure}[h]
\centering
\includegraphics[width=1\textwidth]{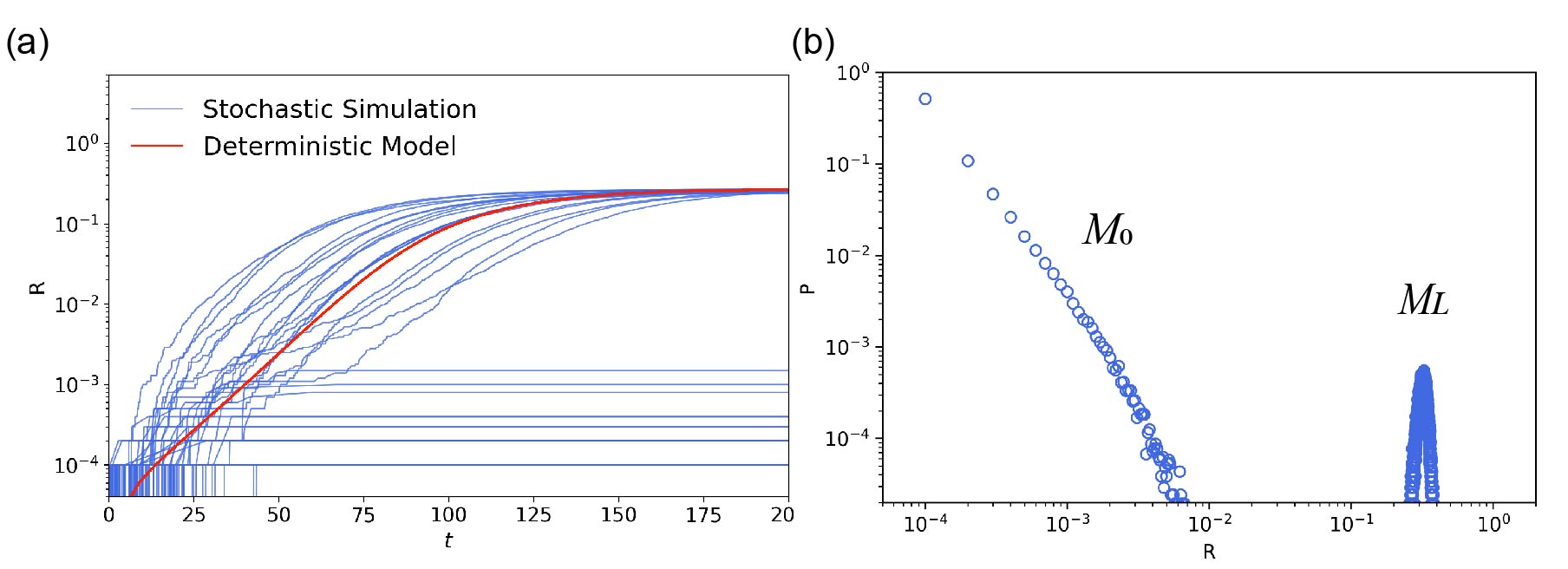}
\caption{
Dynamics of deterministic and stochastic SIR model. (a) Comparing the dynamics of the deterministic SIR model and the trajectories of 200 stochastic simulation samples. (b) Distribution of final recovered population from 100,000 stochastic simulations. Here, ($M_0$) corresponds to stochastic die-out, while ($M_L$) represents stochastic outbreak.}
\label{figure1}
\end{figure}

The bimodal distribution of stochastic take-off and die-out has sparked substantial interest across diverse research domains and applications \cite{lanvcic2011phase,hu2018local,goldenbogen2022control}. Recent studies have centered on utilizing analytical approaches to estimate the probability distribution for stochastic outbreaks and die-outs across various models. For instance, Smith et al. \cite{hindes2022outbreak} rigorously derived outbreak size distributions in stochastic SIR models through master equation analysis, revealing phase transitions between endemic and extinction regimes. Building on this, Parsons et al. \cite{parsons2024probability} quantified epidemic burnout probabilities in demographically structured populations using boundary-layer approximations, demonstrating that even supercritical pathogens (\(R_0 > 1\)) exhibit substantial extinction risks due to finite-population stochasticity.  

However, these analytical methods—while foundational—primarily address what the outbreak or extinction probability is, not whether a specific early-stage transmission will outbreak or die out. A more urgent question arises: Given real-time observations of an emerging contagion, can we predict its fate before the system commits to either bimodal branch? This predictive capability is critical for implementing just-in-time interventions, as the die-out probabilities depend sensitively on initial infectious stages \cite{thompson2016detecting,czuppon2021stochastic,parsons2024probability}. 
Our research addresses this gap by translating theoretical die-out probabilities into actionable predictions. Unlike the master equation formalism in \cite{hindes2022outbreak} or the hybrid ODE-stochastic analysis in \cite{parsons2024probability}, we propose a machine learning framework that directly maps partially observed data (e.g., early infection counts, network topology) to outbreak trajectories. This shift from parameter-dependent probability estimation to data-driven early classification enables proactive resource allocation and real-time emergency response.

\label{}
\section{Definition of stochastic outbreak prediction}
\begin{figure}[h]
\centering
\includegraphics[width=1\textwidth]{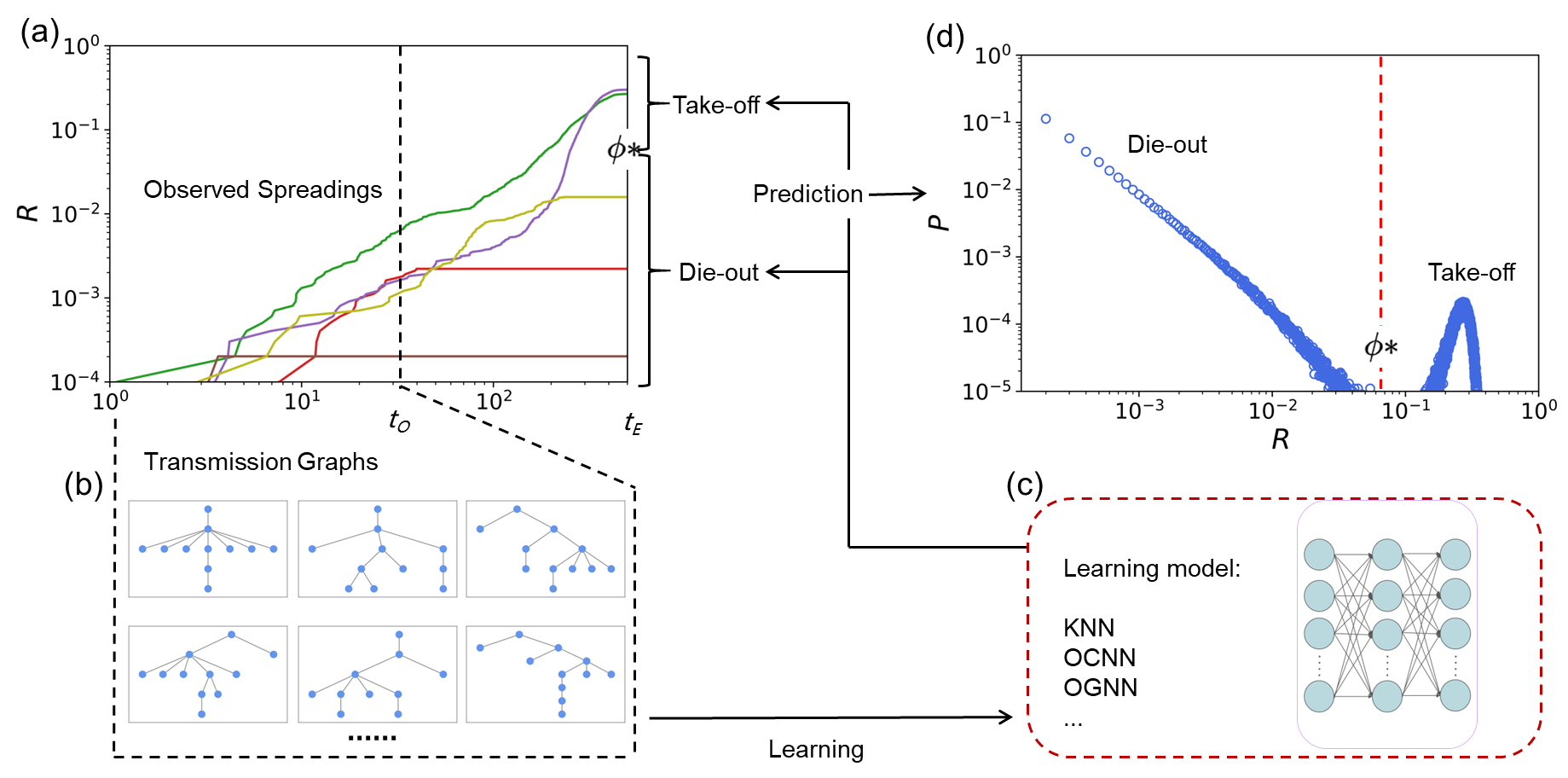}
\caption{
Overview of the stochastic outbreak prediction task. (a) Trajectories of a few stochastic simulation samples. (b) Temporal graphs of several observed transmissions. (c) Prediction method. (d) The distribution of die-outs and take-off results from 100,000 stochastic simulations. \( \phi* \) represents the threshold value for stochastic die-outs and take-off events. \( t_o \) refers to the time observed to predict the outbreak of the spreading. }
\label{prediction_framework}
\end{figure}

In this section, we aim to predict, in the early stage, whether a transmission will stochastically die out or take off. As we described in the previous section, in the early stages of a spreading process, the infection numbers will be few and so random variations alone can cause the spreading to die out or take off. Therefore, predicting in the early stages whether the spread will take off or die out in the future is of paramount importance. Figure \ref{prediction_framework} illustrates our framework for stochastic outbreak prediction. 

More precisely, assume there are a total of \( n \) transmission events  represented as (\( S_1, S_2, \ldots , S_n \)). For each event \( S_i \), we have its transmission dynamic over the time period [0, \( t_E \)], characterized by a series of time moments \( t_k (0 \leq t_k \leq t_E) \) and transmission graph \(\{ G_i(t_0), G_i(t_1), \ldots, G_i(t_{E}) \} \). The time \( t_E \) refers to the ending time of the spreading, and \( t_o \) refers to when we intend to predict the outbreak of the spreading. Let \( n_i \) represent the total number of infections of \( S_i \) at  time \( t_o \). The \( i \)-th transmission sequence, \( S_i \), is defined as a stochastic outbreak  if \( n_i > \phi* \), where \( \phi* \) is the threshold. The vector \( y = (y_1, y_2, \ldots , y_n) \) represents the actual classes of transmission sequences, where each \( y_i \in \{0, 1\} \). In this context, \( y_i = 1 \) if \( S_i \) is a stochastic outbreak,  and \( y_i = 0 \) otherwise.

\begin{equation}
y_i= \begin{cases} 0 & \mathrm{if~}n_i < \phi*,\\1 & \mathrm{if~}n_i \geq \phi*. \end{cases}\
\end{equation}
Let \( \hat{y} = (\hat{y}_1, \hat{y}_2, \ldots, \hat{y}_n) \) be the predicted vector, where each element \( \hat{y}_i \) denotes the outbreak probability of the corresponding spreading,  \( \hat{y}_i \in [0, 1] \). The primary goal of outbreak prediction is to accurately predict the values of \( \hat{y} \). Thus the optimization objective is defined as minimizing the standard binary cross-entropy function: 
\begin{equation}
\mathbf{argmin}-\frac1n\sum_{i=1}^{n}(y_{i}\mathrm{log}(H_{\theta}(S_{i}))+(1-y_{i})\mathrm{log}(1-H_{\theta}(S_{i})))
\end{equation}
Here, \( S_i \) represents the \( i \)th transmission sequence up to \( t_o \). The prediction model, denoted as \( H \), produces an outbreak probability for each input sequence \( S_i \), and \( \theta \) are the model parameters. 

\section{Evaluations Metrics}
As the prediction problem is a binary classification problem, we use the following methods to evaluate the performances of the proposed models:  Accuracy, Precision, Recall, F1-Score, Receiver Operating Characteristic (ROC) curve, and the Area Under the ROC Curve (AUC).  These metrics are calculated by equations (3-6).

\begin{equation}
\\\\Accuracy =\frac{TP+TN}{TP+FP+TN+FN}
\end{equation}

\begin{equation}
\\\\Precision =\frac{TP}{TP+FP}
\end{equation}

\begin{equation}
\\\\Recall=\frac{TP}{TP+FN}
\end{equation}

\begin{equation}
\text{F-Score}=\frac{2\text{TP}}{2\text{TP}+\text{FN}+\text{FP}}
\end{equation}

The definitions of TP, FP, TN, and FN are explained below. 

TP (True Positive): Stochastic take-offs that are properly classified as take-offs.

FP (False Positive): Stochastic die-outs that are wrongly classified as take-offs.

TN (True Negative): Stochastic die-outs that are properly classified as die-outs.

FN (False Negative):  Stochastic take-offs that are wrongly classified as stochastic die-outs.

The ROC curve and AUC measures offer a comprehensive assessment of a classifier's performance by taking into account both sensitivity (the ability to detect true positives) and specificity (the ability to avoid false positives). Therefore, the ROC curve and AUC are commonly considered superior performance metrics for classifiers in scientific research.  The ROC and AUC are measured according to \cite{huang2005using}. 

\section{Prediction methods}

Previous research has indicated that the temporal and structural properties of transmissions both serve as key indicators of spreading virality \cite{meng2018diffusion,sepehr2022structural}. To predict whether an early spreading will stochastically die out or escalate into a significant outbreak, we developed the Outbreak-GWN method capable of learning both the temporal and structural features of early spreadings. 
To evaluate our model's effectiveness, we compare our model with the surveillance thresholds method  \cite{world2023early} and two traditional machine learning methods, namely  KNN and CNN methods. Below, we introduce these methods.

\subsection{Surveillance thresholds }
The surveillance threshold (ST), also referred to as the “early warning threshold”, denotes the minimum number of cases beyond which an infectious disease outbreak is anticipated imminently and necessitates prompt intervention \cite{world2023early}. This method is extensively employed in real-world settings to detect outbreaks of infectious diseases \cite{straetemans2008automatic}. Here we set the ST at 5, 15, and 25 respectively in the experiments.

\subsection{KNN}
K-nearest neighbors (KNN) is a machine learning method that relies on the intuition that similar data points tend to have similar labels \cite{zhang2007ml}. The KNN algorithm classifies a sample based on the labels of its nearest training examples in the feature space. To determine the unknown sample's classification, all distances between the unknown sample and the samples in the training set are computed. The smallest distance value, corresponding to a sample in the training set, is selected as the nearest neighbor for the unknown sample. Consequently, the unknown sample is classified based on its nearest neighbor. The advantages of the KNN model include its robustness, ease of implementation, and its capability to handle preprocessing tasks for large datasets.

\subsection{Outbreak-CNN (OCNN)}
Inspired by the advantages that Convolutional Neural Networks (CNNs) have demonstrated in handling text and image data, we adopted the CNN architecture to learn the temporal patterns of spreading. The architecture of Outbreak-CNN CNN (OCNN) is detailed in Fig.~\ref{appendix:Outbreak-CNN}. Instead of words used in sentence classification, we use transformed infection numbers, derived from the transmission sequences, as inputs to our model. Each input is then transformed into an embedding matrix \( M = [E_1, \ldots, E_n] \), where \( n \) denotes the sequence length. To produce a new feature, a convolution operation is performed using a filter \( w \in \mathbb{R}^{h \times k} \), applied over a window of \( h \) transformed infection numbers. This filter is applied to all possible windows in the sequence, resulting in a feature map \( c \). The feature map is then subjected to a max-over-time pooling operation, where the maximum value \( \hat{c} = \max \{c\} \) is extracted as the feature associated with that specific filter. The objective is to capture the most significant feature, with the highest value, from each feature map. This pooling method effectively handles sequences of varying lengths. Above, we have explained how a feature is acquired from a single filter. The model employs multiple filters (with different window sizes) to obtain various features. These features constitute the penultimate layer and are passed through a fully connected logistic layer, generating an output that represents the probability distribution across labels.

\subsection{Outbreak-GWN}

To capture both the structural and temporal information of spreadings, we developed the Outbreak-GWN (OGWN) model, via the concatenation of the GraphWave method and gated recurrent units neural networks, thus capable of learning both the temporal and structural features of early spreadings. As illustrated in Figure 4, the Outbreak-GWN consists of three main parts: (A) Structural Embedding, (B) Temporal Learning, and (C) Outbreak Prediction. We will delve into them in detail in the following sections. 

\begin{figure}[h]
\centering
\includegraphics[width=1\textwidth]{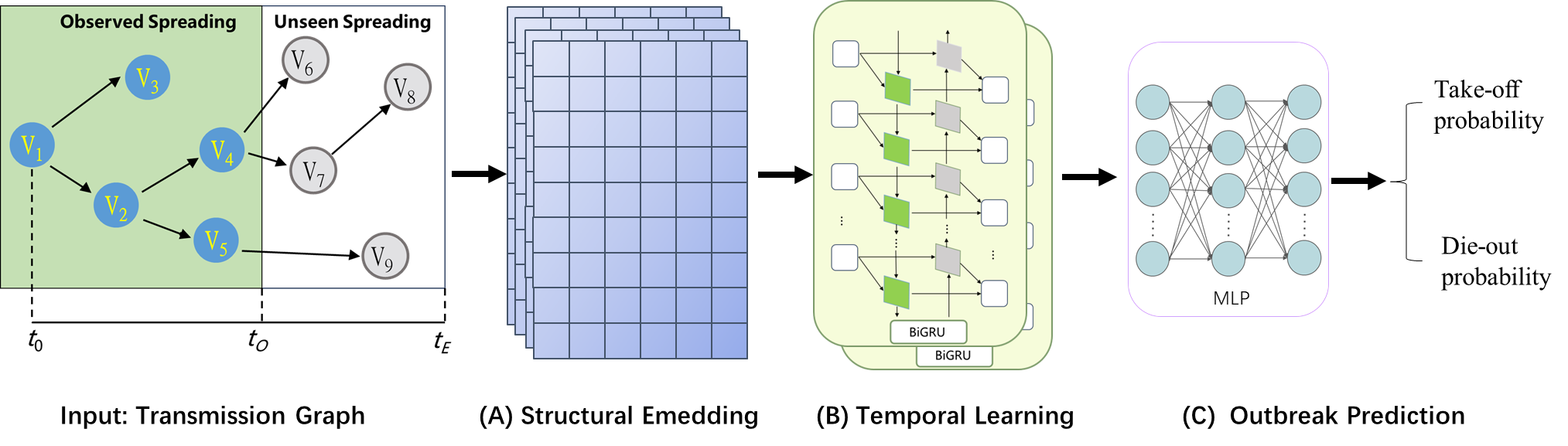}
\caption{The architecture of Outbreak-GWN. }
\label{figure2}
\end{figure}

(A) Structural Embedding. To capture the structural features and obtain a node-level representation, we implement a graph embedding technique that learns the diffusion of a spectral graph wavelet for individual nodes \cite{donnat2018learning}. The GraphWave method leverages the inherent relationship between wavelet coefficients and graph topological properties to recover structurally similar nodes. For a specific node $a$, GraphWave treats its wavelet coefficients as a probability distribution and employs empirical characteristic functions \cite{lukacs1972survey} to represent this distribution. The empirical characteristic function for a scalar random variable $X$ is defined as $\varphi_X(t) = E[e^{itX}]$, where $t \in \mathbb{R}$. Specifically, for a given node $a$ and scale $s$, the empirical characteristic function is defined as:

\begin{equation}
\phi_{a}(t) = \frac{1}{N} \sum_{m=1}^{N} e^{it \psi_{a,m}},
\end{equation}

where $\psi_{a,s}(j)$ denotes the amount of energy that node $a$ has received from node $m$. This approach effectively captures the structural information encoded within the wavelet coefficients, enabling the embedding of nodes with similar topological properties. 

(B) Temporal Learning. To capture the temporal patterns of spreading, we employ the Bidirectional Gated Recurrent Unit (Bi-GRU) \cite{chung2014empirical} to learn the temporal features embedded within the spreading process. GRU is a specific type of recurrent neural network (RNN) that addresses the limitations of traditional RNNs by incorporating gating mechanisms. These mechanisms allow the network to selectively update and reset its hidden state, enabling it to capture long-term dependencies in sequential data more effectively. GRU is calculated using the following formula:

\begin{equation}
\\z_{t}=\sigma(W_{i}*[h_{t-1},x_{t}])\\ 
\end{equation}
\begin{equation}
r_{t}=\sigma(W_{r}*[h_{t-1},x_{t}])\\  
\end{equation}
\begin{equation}
\\ \widetilde{h_{t}}=\mathrm{tanh}(W_{c}*[r_{t}\cdot h_{t-1},x_{t}])\\  
\end{equation}
\begin{equation}
h_{t}=(1-z_{t})\cdot c_{t-1}+z_{t}\cdot\widetilde{h_{t}}\
\end{equation}

where $z_{t}$ is the update gate; $r_{t}$ is the reset gate; $\widetilde{h_{t}}$ is the candidate hidden state; $h_{t}$ is the hidden state; $\sigma$ denote the activation function; $x_t$ denote the input at t time; $W_i,W_r,W_c$ are the corresponding weight matrixes of $z_t$, $r_t$, $h_t$. $h_{t-1}$ is the hidden state at $t-1$ time. 

 The Bi-GRU architecture utilizes two GRUs: a forward GRU ($GRU_{fwd}$) that reads the sequence from left to right, and a backward GRU ($GRU_{bwd}$) that reads from right to left. By concatenating the outputs of the $GRU_{fwd}$) and $GRU_{bwd}$), the final representation $\overleftrightarrow{h}_{t}$ can be obtained as:
 \begin{equation}
 \begin{aligned}
&\overrightarrow{h}_{t} =\mathrm{GRU}_{fwd}(x_{t}, \overrightarrow{h}_{t-1}), \\
&\overleftarrow{h}_{t} =\mathrm{GRU}_{bwd}(x_{t}, \overleftarrow{h}_{t+1}), \\
&\overleftrightarrow{h}_{t} =\overrightarrow{h}_t\oplus\overleftarrow{h}_t. 
\end{aligned}
\end{equation}
 This concatenating operation enables the Bi-GRU to effectively learn long-term dependencies within sequences by capturing information from both temporal directions.

(C) Prediction. The output $\overleftrightarrow{h}_{t}$ from the BiGRU layer is fed into the Multi-Layer Perceptron (MLP) layer to get the final prediction as: 

\begin{equation}
\\ H_{\theta}(S_{i}) = MLP(\overleftrightarrow{h}_{t})
\end{equation}

\section{Model performance}

In this section, we present the results of stochastic outbreak prediction in two classic network models: the Erdős–Rényi (ER) network and the Barabási–Albert (BA) network. To evaluate the robustness and generalizability of these methods across various levels of infectiousness, we conducted experiments in scenarios characterized by low, medium, and high infectivity levels, respectively.

\subsection{Prediction of take-offs in the BA Network}

\begin{figure}[h]
\centering
\includegraphics[width=1\textwidth]{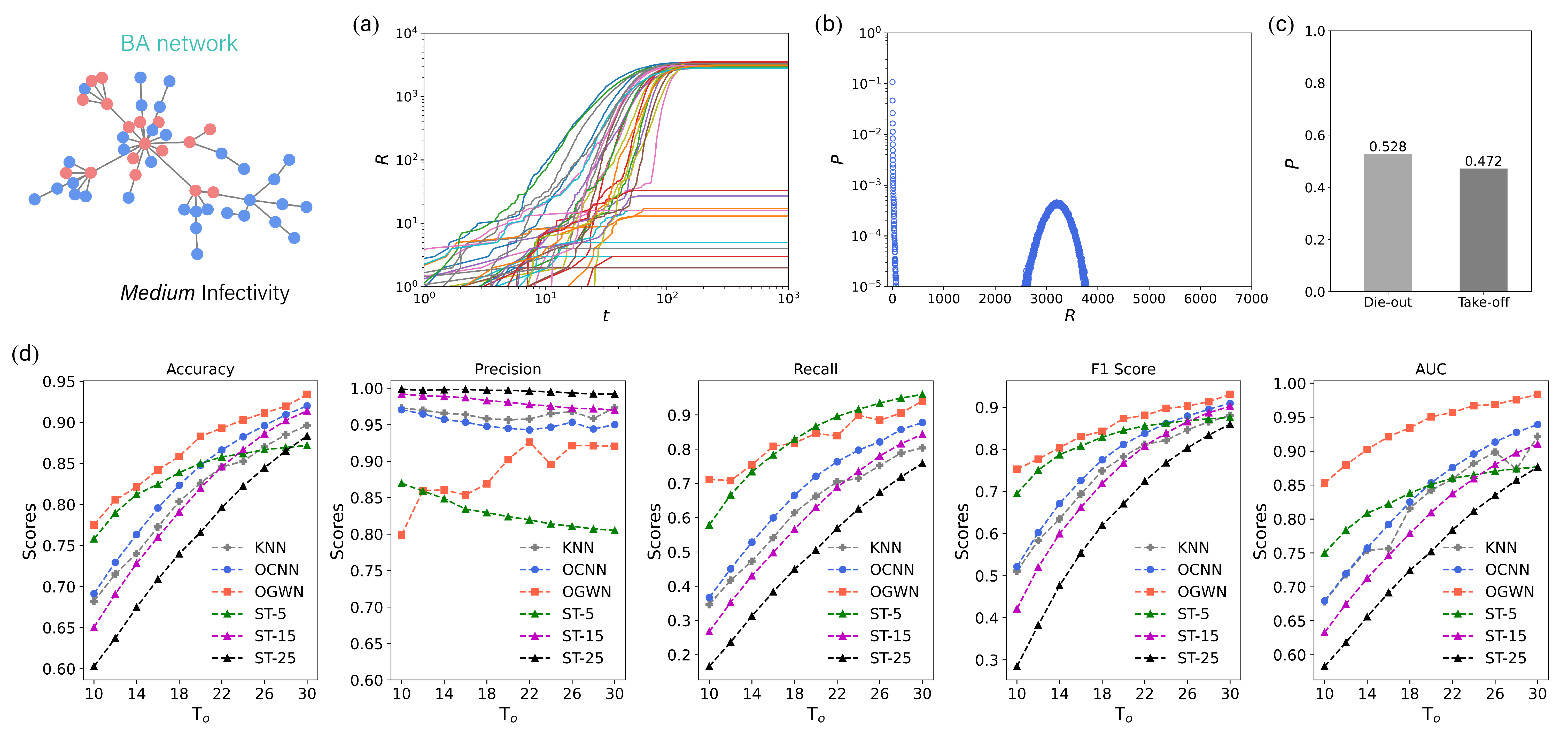}
\caption{Stochastic outbreak prediction with medium infectivity scenario in BA network. (a) 100 trajectories simulated with medium infectivity $\beta = 0.02$. (b) The distribution of final recovers of 500000 stochastic simulations with medium infectivity $\beta = 0.02$. (c) Probability of stochastic die-out and take-off. (d) Model performances across varying observation times  \( T_o \) ranging from 10 to 30.  }
\label{fig_medium_infectivity}
\end{figure}

We first examine the outbreak prediction in a BA network, which consists of N = 10,000 nodes and is characterized by the parameter \textit{m} = 3. To investigate the predictions under low, medium, and high infectivity conditions, we selected transmission rates of $\beta$ = 0.015, 0.02, and 0.03, resulting in average outbreak sizes affecting 19\%, 32\%, and 53\% of the network, respectively. We present the detailed spreading information and model performance for the medium infectivity scenario in Fig. \ref{fig_medium_infectivity}.

To compare the performance of our proposed model with other approaches, we employed the metrics of  Accuracy, Recall, Precision, F1-Score, and AUC to evaluate the models across varying observation times. As shown in Fig.\ref{fig_medium_infectivity}(d), the Outbreak-GWN model consistently demonstrated superior performance in terms of Accuracy, F1-Score, and AUC across all observation times. With respect to the Precision and Recall metrics, higher values for Precision were typically associated with lower values for Recall, and vice versa, which reflects the well-established Precision-Recall trade-off as extensively discussed in previous literature \cite{buckland1994relationship, gordon1989recall}.

To evaluate the robustness and generalizability of the proposed models across various diseases, experiments were then conducted on scenarios representing low and high infectivity levels, respectively. Further analysis of scenarios with low and high infectivity levels  (see Fig.\ref{fig:BA_low_infectivity} and \ref{fig:BA_high_infectivity})  also revealed that the Outbreak-GWN model outperformed other methods across most observation intervals, confirming its robustness in predicting outbreaks under different infectivity conditions. 

Notably, the results also indicate that the ST method exhibited a high degree of performance instability across varying infectivity conditions.  For instance, while ST-5 generally performed well in high infectivity scenarios, its performance significantly declined in low infectivity scenarios. 

\subsection{Prediction of Outbreaks in the ER Network}

\begin{figure}[h]
\centering
\includegraphics[width=1\textwidth]{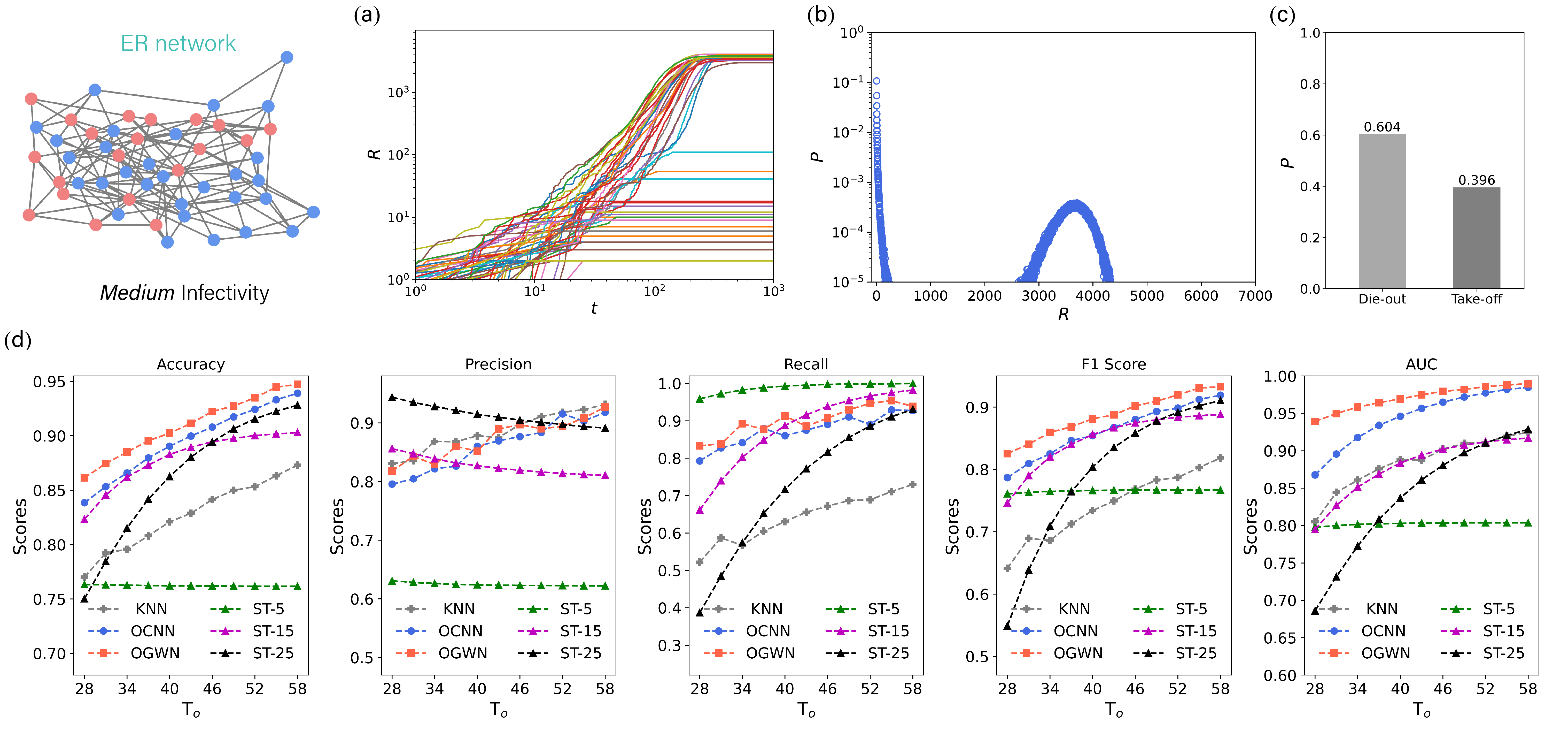}
\caption{Stochastic outbreak prediction with medium infectivity scenario in ER network. (a) 100 trajectories simulated with medium infectivity $\beta = 0.033$. (b)  The distribution of final recoveries of 500000 stochastic simulations with medium infectivity $\beta = 0.033$. (c) Probability of stochastic die-out and take-off. (d) Model performances across varying observation times  \( T_o \) ranging from 28 to 58.  }
\label{fig_medium_infectivity_er}
\end{figure}

In this section, we examine the outbreak prediction in the ER network, which consists of N = 10,000 nodes with an average degree of \textit{k} = 5. To investigate the prediction performance under low, medium, and high infectivity conditions, we selected transmission rates of $\beta$ = 0.03, 0.033, and 0.04, resulting in average outbreak sizes affecting 28\%, 38\%, and 55\% of the network, respectively. Figure\ref{fig_medium_infectivity_er} illustrates the detailed spreading information and model performance for the medium infectivity scenario of spreading in the ER network.

As shown in Fig.\ref{fig_medium_infectivity_er}(d), the Outbreak-GWN model consistently exhibited superior performance across all observation times, as measured by Accuracy, F1-Score, and AUC metrics.  Further analysis of scenarios with low and high infectivity levels (see Fig.\ref{fig:ER_low_infectivity_er} and \ref{fig:ER_high_infectivity_er}) also shows that the Outbreak-GWN model outperformed other methods across most observation intervals. The results obtained from the ER network scenario, combined with those from the BA network scenario, confirm the robustness and generalizability of the Outbreak-GWN model in predicting stochastic outbreaks across various infectivity conditions and different network models. 

Interestingly, we observed that the ST method exhibited significant performance instability between the prediction task in the BA and ER network models.  For example, while ST-5 generally performed well in outbreak predictions on the BA network, it showed the poorest performance on the ER network compared to the ST-15 and ST-25 methods. This finding further underscores the robustness and generalizability of our proposed Outbreak-GWN model.

\section{Pretrain-Finetune Framework for Outbreak Prediction}

Directly training neural networks for epidemic spreading in scenarios with limited training data, such as new emerging infectious diseases or spreading processes in novel environments  (such as novel contact networks), is often infeasible. These challenges are exacerbated by the inherent difficulties in acquiring real-world epidemiological data, including high collection costs, incompleteness, and inaccuracies stemming from the complexities of multi-setting surveillance.

\begin{figure}[h]
\centering
\includegraphics[width=1\textwidth]{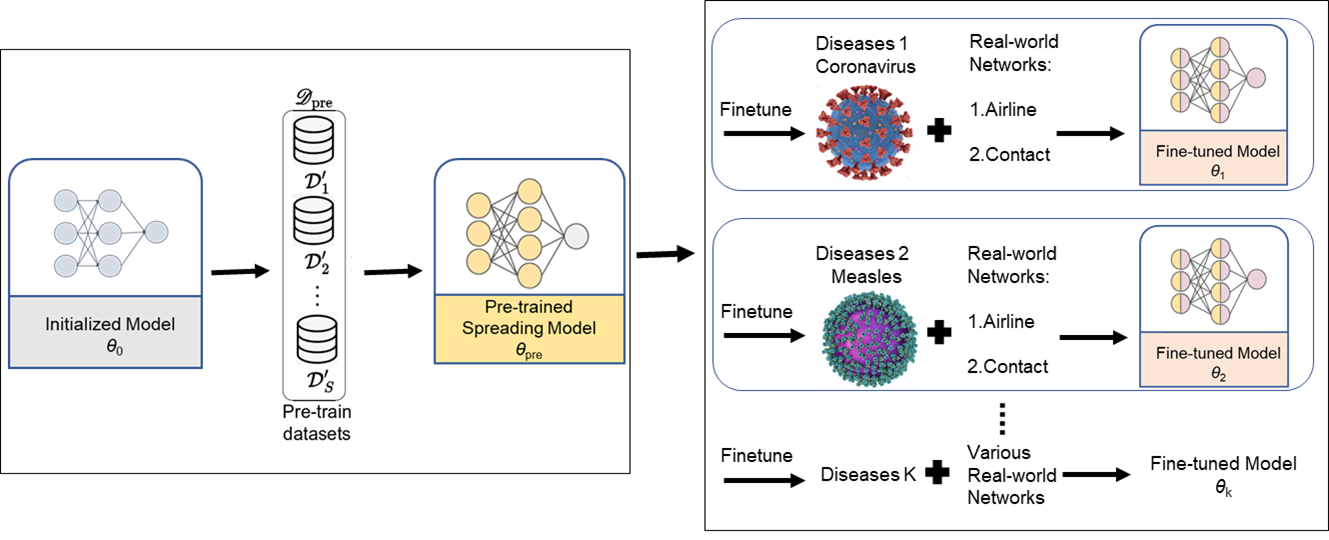}
\caption{The architecture of our pretrain-finetune model }
\label{pretrain_finetune model}
\end{figure}

To overcome these limitations, we propose a transfer learning strategy leveraging a pretrain-finetune framework with simulation data.  Our approach involves first pretraining a neural network on a diverse set of in silico outbreak simulations spanning a wide epidemiological parameter space. Subsequently, the pretrained model is fine-tuned on smaller, scenario-specific datasets ( demonstrated in Fig.\ref{pretrain_finetune model}). This allows the model to adapt to new contexts while leveraging learned fundamental disease transmission dynamics. Critically, to ensure cross-network transferability, fine-tuning data is exclusively derived from networks separate from the pretraining datasets, guaranteeing robust performance in completely unseen scenarios. The model also demonstrates robust capability to predict stochastic outbreaks across diverse infectivity conditions and network topologies, highlighting its real-world applicability for both emerging and re-emerging pathogens.
We validate this pretrain-finetune framework through case studies involving COVID-19 and measles in two distinct networks: an airline travel network and an empirical social contact network.

\subsection{Implementation Details and Experimental Results}

To maximize generalizability, we pretrain the model using simulated outbreaks generated across diverse epidemiological conditions. We systematically vary infection parameters on reference networks to generate scenarios spanning subcritical, critical, and supercritical regimes. Through self-supervised graph contrastive learning, the model learns universal spreading pattern representations from this simulation data during pretraining, before adapting to specific networks and epidemics through targeted fine-tuning.
This pretraining phase employs contrastive learning to distill robust transmission patterns, resulting in a model capable of effectively initializing predictions in novel scenarios.

For fine-tuning, we train the pretrained model using limited data from COVID-19 and measles outbreaks, employing an attention-based architecture.  This stage requires significantly fewer epochs than pretraining (e.g., 10 epochs vs. 100 during pretraining), leveraging the pretrained model's learned prior knowledge. 

As shown in Table \ref{table1}, our framework consistently outperforms baseline models, including OGWN, in predicting novel outbreaks, achieving superior accuracy despite training on minimal scenario-specific data. Notably, as shown in Table \ref{table1}, our Pretrain-Finetune model achieves improved AUC scores compared to OGWN across various datasets and observation time windows. Quantitatively, the “Improvement” row shows the percentage increase in performance of Pretrain-Finetune compared to OGWN. We observe a significant boost, consistently demonstrating gains such as a +5.3\% improvement on the Measles-G8 dataset with an observation time window of 5 (OBT-5), and +4.0\% on the Measles-Airline dataset (OBT-5).  The improvements are consistently positive across all datasets and time windows, further solidifying the robustness of the proposed approach. This performance advantage underscores the strategic benefit of combining simulation-based pretraining with targeted fine-tuning—a paradigm particularly suited for early-stage epidemics where real-world data is scarce.

\begin{table}[htbp]
\caption{Model Performance Comparison with Pretrain-Finetune vs other models}
\label{tab:transposed_auc}
\centering
\resizebox{\textwidth}{!}{%
\begin{tabular}{l *{4}{rr}}
\toprule
\multirow{2}{*}{Model} & 
\multicolumn{2}{c}{Measles-G8} & 
\multicolumn{2}{c}{Measles-Airline} & 
\multicolumn{2}{c}{Covid19-Airline} & 
\multicolumn{2}{c}{Covid19-G8} \\
\cmidrule(lr){2-3} \cmidrule(lr){4-5} \cmidrule(lr){6-7} \cmidrule(lr){8-9}
& OBT-5 & OBT-10 & OBT-5 & OBT-10 & OBT-10 & OBT-20 & OBT-10 & OBT-20 \\
\midrule
ST-5              & 0.799 & 0.804 & 0.696 & 0.832 & 0.764 & 0.887 & 0.757 & 0.785 \\
ST-15             & 0.790 & 0.924 & 0.539 & 0.707 & 0.596 & 0.799 & 0.686 & 0.852 \\
ST-25             & 0.736 & 0.915 & 0.511 & 0.644 & 0.541 & 0.738 & 0.616 & 0.812 \\
KNN               & 0.781 & 0.913 & 0.500 & 0.749 & 0.608 & 0.814 & 0.697 & 0.837 \\
OCNN              & 0.820 & 0.952 & 0.583 & 0.750 & 0.655 & 0.845 & 0.748 & 0.905 \\
OGWN              & 0.862 & 0.963 & 0.807 & 0.885 & 0.834 & 0.924 & 0.832 & 0.933 \\
Pretrain-Finetune & $\mathbf{0.908}$ & $\mathbf{0.980}$ & $\mathbf{0.839}$ & $\mathbf{0.916}$ & $\mathbf{0.871}$ & $\mathbf{0.952}$ & $\mathbf{0.851}$ & $\mathbf{0.943}$ \\
\addlinespace
Improvement       & (+5.3\%) & (+1.8\%) & (+4.0\%) & (+3.5\%) & (+4.4\%) & (+3.0\%) & (+2.3\%) & (+1.1\%) \\
\bottomrule
\end{tabular}%
}
\begin{tablenotes}
\small
\item Improvement calculated as: $\frac{\text{Finetune} - \text{OGWN}}{\text{OGWN}} \times 100\%$
\end{tablenotes}

\label{table1}
\end{table}

\section{Conclusion and Discussion}

The integration of epidemic modeling with artificial intelligence represents a paradigm shift in addressing the dynamic challenges of disease modeling \cite{rodriguez2024machine,kraemer2025artificial}. While deterministic models have historically dominated epidemiological frameworks \cite{kermack1927contribution, pastor2015epidemic}, their inability to account for the inherent stochasticity of transmission events limits their realism and predictive accuracy. Stochastic modeling has emerged as a powerful tool for capturing the intrinsic randomness and complexity of natural phenomena, particularly in infectious disease dynamics. Individual-based stochastic epidemic models have demonstrated their effectiveness in representing the nuanced dynamics of outbreaks and informing policy decisions \cite{lloyd2005superspreading, allen2015stochastic, czuppon2021stochastic}. This study builds on these foundations by integrating stochastic epidemic modeling with advanced machine learning techniques, offering a novel approach to understanding and predicting early-stage outbreak dynamics.

By incorporating network heterogeneity and the stochastic nature of transmission processes, we developed a deep-learning framework capable of predicting stochastic take-offs and die-outs with high accuracy. The proposed Outbreak-GWN model leverages both structural and temporal information to forecast the early stages of outbreaks, demonstrating robustness and generalizability across diverse infectivity scenarios and network topologies, including Erdős–Rényi (ER) and Barabási–Albert (BA) networks. This framework represents a significant advancement in the field, as it combines stochastic epidemic modeling with machine learning for the first time, offering a comprehensive tool for predicting and mitigating the impact of infectious diseases.

The ability to distinguish between stochastic die-outs and impending take-offs is critical for deploying timely and targeted interventions. Timely identification of emerging outbreaks facilitates rapid public health response, which is essential for curbing further transmission and minimizing epidemiological consequences \cite{stone2007seasonal}.  
Whether the goal is to contain a pathogen before it escalates into a major epidemic or to curb the spread of misinformation before it goes viral, the adaptability of our framework to different network structures and infection scenarios suggests broad applicability beyond epidemiology. Potential extensions include predicting innovation diffusion or managing information cascades in social networks.
Furthermore, the integration of physics-based modeling with machine learning, as highlighted in recent literature \cite{carleo2019machine, hofman2021integrating, karniadakis2021physics,ye2025integrating}, has proven instrumental in enhancing the reliability and utility of predictive models. Our work aligns with this trend, demonstrating how hybrid approaches can improve the understanding and prediction of early-stage outbreak and contribute to more effective public health strategies.
Ultimately, this study underscores the potential of combining stochastic modeling with machine learning to address complex challenges in epidemiology and beyond, offering a pathway to mitigate the impact of emerging infectious diseases and other dynamic processes.

\label{}

\section*{Code and data availability}
To ensure reproducibility and encourage further research, all implementation code and data will be made publicly available on GitHub upon publication. 

\clearpage

\biboptions{numbers,sort&compress}
\bibliographystyle{elsarticle-num} 
\bibliography{ref}

\begin{thebibliography}{10}
\expandafter\ifx\csname url\endcsname\relax
  \def\url#1{\texttt{#1}}\fi
\expandafter\ifx\csname urlprefix\endcsname\relax\def\urlprefix{URL }\fi
\expandafter\ifx\csname href\endcsname\relax
  \def\href#1#2{#2} \def\path#1{#1}\fi

\bibitem{jackson2007diffusion}
M.~O. Jackson, L.~Yariv, Diffusion of behavior and equilibrium properties in network games, American Economic Review 97~(2) (2007) 92--98.

\bibitem{ugander2012structural}
J.~Ugander, L.~Backstrom, C.~Marlow, J.~Kleinberg, Structural diversity in social contagion, Proceedings of the national academy of sciences 109~(16) (2012) 5962--5966.

\bibitem{cheng2014can}
J.~Cheng, L.~Adamic, P.~A. Dow, J.~M. Kleinberg, J.~Leskovec, Can cascades be predicted?, in: Proceedings of the 23rd international conference on World wide web, 2014, pp. 925--936.

\bibitem{rosenthal2015revealing}
S.~B. Rosenthal, C.~R. Twomey, A.~T. Hartnett, H.~S. Wu, I.~D. Couzin, Revealing the hidden networks of interaction in mobile animal groups allows prediction of complex behavioral contagion, Proceedings of the National Academy of Sciences 112~(15) (2015) 4690--4695.

\bibitem{scarpino2019predictability}
S.~V. Scarpino, G.~Petri, On the predictability of infectious disease outbreaks, Nature communications 10~(1) (2019) 898.

\bibitem{estrada2020covid}
E.~Estrada, Covid-19 and sars-cov-2. modeling the present, looking at the future, Physics Reports 869 (2020) 1--51.

\bibitem{salganik2006experimental}
M.~J. Salganik, P.~S. Dodds, D.~J. Watts, Experimental study of inequality and unpredictability in an artificial cultural market, science 311~(5762) (2006) 854--856.

\bibitem{bertozzi2020challenges}
A.~L. Bertozzi, E.~Franco, G.~Mohler, M.~B. Short, D.~Sledge, The challenges of modeling and forecasting the spread of covid-19, Proceedings of the National Academy of Sciences 117~(29) (2020) 16732--16738.

\bibitem{rosenkrantz2022fundamental}
D.~J. Rosenkrantz, A.~Vullikanti, S.~Ravi, R.~E. Stearns, S.~Levin, H.~V. Poor, M.~V. Marathe, Fundamental limitations on efficiently forecasting certain epidemic measures in network models, Proceedings of the National Academy of Sciences 119~(4) (2022) e2109228119.

\bibitem{castro2020turning}
M.~Castro, S.~Ares, J.~A. Cuesta, S.~Manrubia, The turning point and end of an expanding epidemic cannot be precisely forecast, Proceedings of the National Academy of Sciences 117~(42) (2020) 26190--26196.

\bibitem{wilke2020predicting}
C.~O. Wilke, C.~T. Bergstrom, Predicting an epidemic trajectory is difficult, Proceedings of the National Academy of Sciences 117~(46) (2020) 28549--28551.

\bibitem{der2009aleatory}
A.~Der~Kiureghian, O.~Ditlevsen, Aleatory or epistemic? does it matter?, Structural safety 31~(2) (2009) 105--112.

\bibitem{penn2023intrinsic}
M.~J. Penn, D.~J. Laydon, J.~Penn, C.~Whittaker, C.~Morgenstern, O.~Ratmann, S.~Mishra, M.~S. Pakkanen, C.~A. Donnelly, S.~Bhatt, Intrinsic randomness in epidemic modelling beyond statistical uncertainty, Communications Physics 6~(1) (2023) 146.

\bibitem{kermack1927contribution}
W.~O. Kermack, A.~G. McKendrick, A contribution to the mathematical theory of epidemics, Proceedings of the royal society of london. Series A, Containing papers of a mathematical and physical character 115~(772) (1927) 700--721.

\bibitem{pastor2015epidemic}
R.~Pastor-Satorras, C.~Castellano, P.~Van~Mieghem, A.~Vespignani, Epidemic processes in complex networks, Reviews of modern physics 87~(3) (2015) 925.

\bibitem{cai2022modeling}
J.~Cai, X.~Deng, J.~Yang, K.~Sun, H.~Liu, Z.~Chen, C.~Peng, X.~Chen, Q.~Wu, J.~Zou, et~al., Modeling transmission of sars-cov-2 omicron in china, Nature medicine 28~(7) (2022) 1468--1475.

\bibitem{czuppon2021stochastic}
P.~Czuppon, E.~Schertzer, F.~Blanquart, F.~D{\'e}barre, The stochastic dynamics of early epidemics: probability of establishment, initial growth rate, and infection cluster size at first detection, Journal of The Royal Society Interface 18~(184) (2021) 20210575.

\bibitem{britton2019stochastic}
T.~Britton, E.~Pardoux, F.~Ball, C.~Laredo, D.~Sirl, V.~C. Tran, Stochastic epidemic models with inference, Vol. 2255, Springer, 2019.

\bibitem{thompson2016detecting}
R.~N. Thompson, C.~A. Gilligan, N.~J. Cunniffe, Detecting presymptomatic infection is necessary to forecast major epidemics in the earliest stages of infectious disease outbreaks, PLoS computational biology 12~(4) (2016) e1004836.

\bibitem{lloyd2005superspreading}
J.~O. Lloyd-Smith, S.~J. Schreiber, P.~E. Kopp, W.~M. Getz, Superspreading and the effect of individual variation on disease emergence, Nature 438~(7066) (2005) 355--359.

\bibitem{lemieux2021phylogenetic}
J.~E. Lemieux, K.~J. Siddle, B.~M. Shaw, C.~Loreth, S.~F. Schaffner, A.~Gladden-Young, G.~Adams, T.~Fink, C.~H. Tomkins-Tinch, L.~A. Krasilnikova, et~al., Phylogenetic analysis of sars-cov-2 in boston highlights the impact of superspreading events, Science 371~(6529) (2021) eabe3261.

\bibitem{sun2021transmission}
K.~Sun, W.~Wang, L.~Gao, Y.~Wang, K.~Luo, L.~Ren, Z.~Zhan, X.~Chen, S.~Zhao, Y.~Huang, et~al., Transmission heterogeneities, kinetics, and controllability of sars-cov-2, Science 371~(6526) (2021) eabe2424.

\bibitem{sneppen2021overdispersion}
K.~Sneppen, B.~F. Nielsen, R.~J. Taylor, L.~Simonsen, Overdispersion in covid-19 increases the effectiveness of limiting nonrepetitive contacts for transmission control, Proceedings of the National Academy of Sciences 118~(14) (2021) e2016623118.

\bibitem{dangerfield2009integrating}
C.~Dangerfield, J.~V. Ross, M.~J. Keeling, Integrating stochasticity and network structure into an epidemic model, Journal of the Royal Society Interface 6~(38) (2009) 761--774.

\bibitem{andersson2012stochastic}
H.~Andersson, T.~Britton, Stochastic epidemic models and their statistical analysis, Vol. 151, Springer Science \& Business Media, 2012.

\bibitem{gillespie1977exact}
D.~T. Gillespie, Exact stochastic simulation of coupled chemical reactions, The journal of physical chemistry 81~(25) (1977) 2340--2361.

\bibitem{allen2015stochastic}
L.~J. Allen, Stochastic population and epidemic models, Mathematical biosciences lecture series, stochastics in biological systems 1 (2015) 120--128.

\bibitem{colizza2006role}
V.~Colizza, A.~Barrat, M.~Barth{\'e}lemy, A.~Vespignani, The role of the airline transportation network in the prediction and predictability of global epidemics, Proceedings of the National Academy of Sciences 103~(7) (2006) 2015--2020.

\bibitem{geng2021kernel}
X.~Geng, G.~G. Katul, F.~Gerges, E.~Bou-Zeid, H.~Nassif, M.~C. Boufadel, A kernel-modulated sir model for covid-19 contagious spread from county to continent, Proceedings of the National Academy of Sciences 118~(21) (2021) e2023321118.

\bibitem{lanvcic2011phase}
A.~Lan{\v{c}}i{\'c}, N.~Antulov-Fantulin, M.~{\v{S}}iki{\'c}, H.~{\v{S}}tefan{\v{c}}i{\'c}, Phase diagram of epidemic spreading—unimodal vs. bimodal probability distributions, Physica A: Statistical Mechanics and its Applications 390~(1) (2011) 65--76.

\bibitem{hu2018local}
Y.~Hu, S.~Ji, Y.~Jin, L.~Feng, H.~E. Stanley, S.~Havlin, Local structure can identify and quantify influential global spreaders in large scale social networks, Proceedings of the National Academy of Sciences 115~(29) (2018) 7468--7472.

\bibitem{goldenbogen2022control}
B.~Goldenbogen, S.~O. Adler, O.~Bodeit, J.~A. Wodke, X.~Escalera-Fanjul, A.~Korman, M.~Krantz, L.~Bonn, R.~Mor{\'a}n-Torres, J.~E. Haffner, et~al., Control of covid-19 outbreaks under stochastic community dynamics, bimodality, or limited vaccination, Advanced Science 9~(23) (2022) 2200088.

\bibitem{hindes2022outbreak}
J.~Hindes, M.~Assaf, I.~B. Schwartz, Outbreak size distribution in stochastic epidemic models, Physical Review Letters 128~(7) (2022) 078301.

\bibitem{parsons2024probability}
T.~L. Parsons, B.~M. Bolker, J.~Dushoff, D.~J. Earn, The probability of epidemic burnout in the stochastic sir model with vital dynamics, Proceedings of the National Academy of Sciences 121~(5) (2024) e2313708120.

\bibitem{huang2005using}
J.~Huang, C.~X. Ling, Using auc and accuracy in evaluating learning algorithms, IEEE Transactions on knowledge and Data Engineering 17~(3) (2005) 299--310.

\bibitem{meng2018diffusion}
J.~Meng, W.~Peng, P.-N. Tan, W.~Liu, Y.~Cheng, A.~Bae, Diffusion size and structural virality: The effects of message and network features on spreading health information on twitter, Computers in human behavior 89 (2018) 111--120.

\bibitem{sepehr2022structural}
A.~Sepehr, H.~Beigy, Structural virality estimation and maximization in diffusion networks, Expert Systems with Applications 206 (2022) 117657.

\bibitem{world2023early}
W.~H. Organization, et~al., Early warning alert and response in emergencies: an operational guide, World Health Organization, 2023.

\bibitem{straetemans2008automatic}
M.~Straetemans, D.~Altmann, T.~Eckmanns, G.~Krause, Automatic outbreak detection algorithm versus electronic reporting system, Emerging infectious diseases 14~(10) (2008) 1610.

\bibitem{zhang2007ml}
M.-L. Zhang, Z.-H. Zhou, Ml-knn: A lazy learning approach to multi-label learning, Pattern recognition 40~(7) (2007) 2038--2048.

\bibitem{donnat2018learning}
C.~Donnat, M.~Zitnik, D.~Hallac, J.~Leskovec, Learning structural node embeddings via diffusion wavelets, in: Proceedings of the 24th ACM SIGKDD international conference on knowledge discovery \& data mining, 2018, pp. 1320--1329.

\bibitem{lukacs1972survey}
E.~Lukacs, A survey of the theory of characteristic functions, Advances in Applied Probability 4~(1) (1972) 1--37.

\bibitem{chung2014empirical}
J.~Chung, C.~Gulcehre, K.~Cho, Y.~Bengio, Empirical evaluation of gated recurrent neural networks on sequence modeling, arXiv preprint arXiv:1412.3555 (2014).

\bibitem{buckland1994relationship}
M.~Buckland, F.~Gey, The relationship between recall and precision, Journal of the American society for information science 45~(1) (1994) 12--19.

\bibitem{gordon1989recall}
M.~Gordon, M.~Kochen, Recall-precision trade-off: A derivation, Journal of the American Society for Information Science 40~(3) (1989) 145--151.

\bibitem{rodriguez2024machine}
A.~Rodriguez, H.~Kamarthi, P.~Agarwal, J.~Ho, M.~Patel, S.~Sapre, B.~A. Prakash, Machine learning for data-centric epidemic forecasting, Nature Machine Intelligence 6~(10) (2024) 1122--1131.

\bibitem{kraemer2025artificial}
M.~U. Kraemer, J.~L.-H. Tsui, S.~Y. Chang, S.~Lytras, M.~P. Khurana, S.~Vanderslott, S.~Bajaj, N.~Scheidwasser, J.~L. Curran-Sebastian, E.~Semenova, et~al., Artificial intelligence for modelling infectious disease epidemics, Nature 638~(8051) (2025) 623--635.

\bibitem{stone2007seasonal}
L.~Stone, R.~Olinky, A.~Huppert, Seasonal dynamics of recurrent epidemics, Nature 446~(7135) (2007) 533--536.

\bibitem{carleo2019machine}
G.~Carleo, I.~Cirac, K.~Cranmer, L.~Daudet, M.~Schuld, N.~Tishby, L.~Vogt-Maranto, L.~Zdeborov{\'a}, Machine learning and the physical sciences, Reviews of Modern Physics 91~(4) (2019) 045002.

\bibitem{hofman2021integrating}
J.~M. Hofman, D.~J. Watts, S.~Athey, F.~Garip, T.~L. Griffiths, J.~Kleinberg, H.~Margetts, S.~Mullainathan, M.~J. Salganik, S.~Vazire, et~al., Integrating explanation and prediction in computational social science, Nature 595~(7866) (2021) 181--188.

\bibitem{karniadakis2021physics}
G.~E. Karniadakis, I.~G. Kevrekidis, L.~Lu, P.~Perdikaris, S.~Wang, L.~Yang, Physics-informed machine learning, Nature Reviews Physics 3~(6) (2021) 422--440.

\bibitem{ye2025integrating}
Y.~Ye, A.~Pandey, C.~Bawden, D.~M. Sumsuzzman, R.~Rajput, A.~Shoukat, B.~H. Singer, S.~M. Moghadas, A.~P. Galvani, Integrating artificial intelligence with mechanistic epidemiological modeling: a scoping review of opportunities and challenges, Nature Communications 16~(1) (2025) 581.

\end{thebibliography}






\section{Acknowledgements }

This research was supported by the Fundamental Research Funds for the Central Universities (No. SWU-XDJH202303), and the Natural Science Foundation of China (No. 72374173).

\section{Author contributions statement}

T.J., and W.H. conceived the research. W.H. conducted the research and drafted the manuscript. All authors reviewed and approved the final version.

\section{Ethics declarations}

The authors declare no competing interests.

\clearpage

\appendix

\section{The architecture of Outbreak-CNN}
\label{appendix:Outbreak-CNN}

\begin{figure*}[h]
\centering
\includegraphics[scale=0.3]{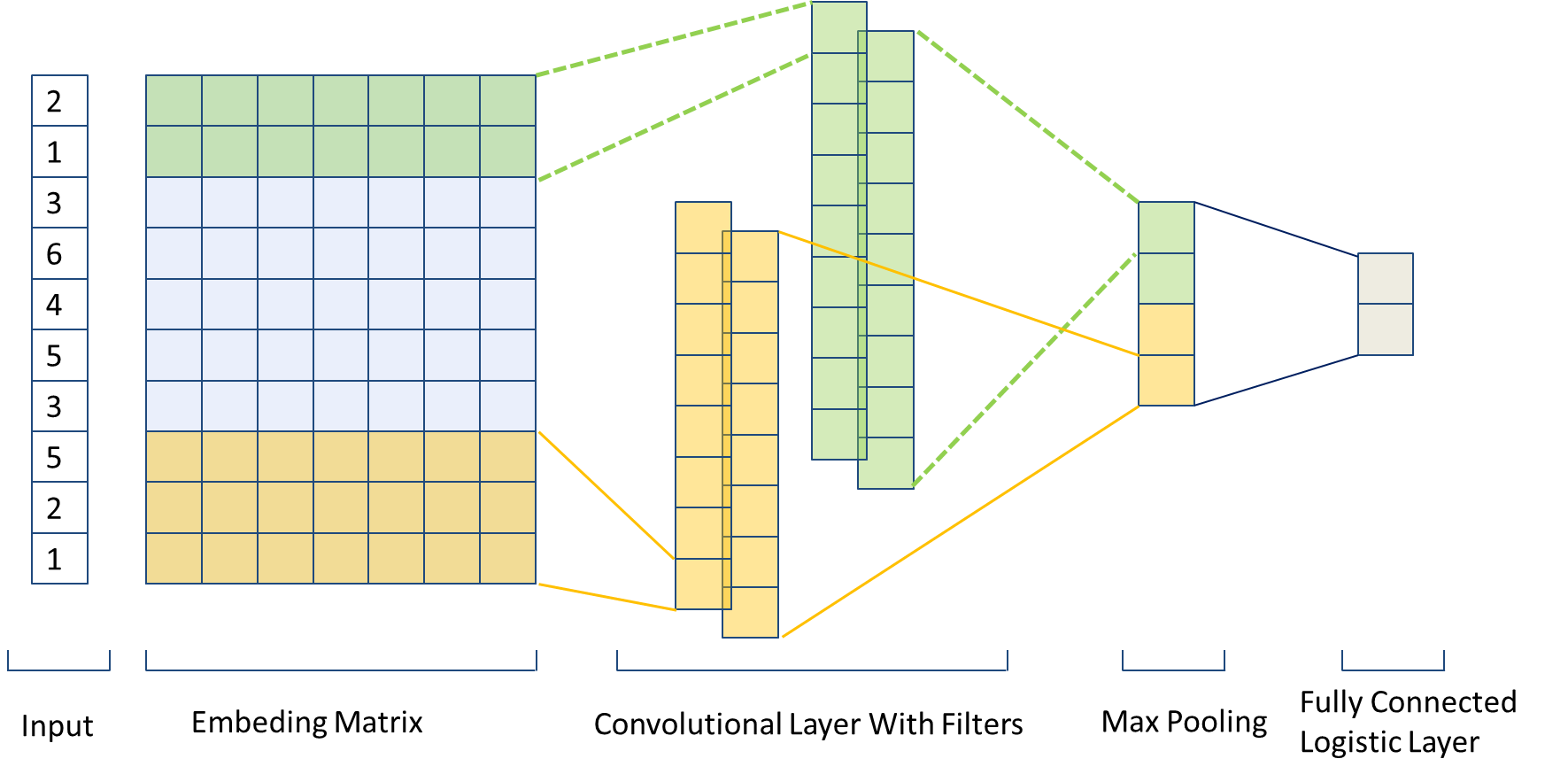}
\caption{The architecture of Outbreak-CNN}
\end{figure*}

\clearpage

\section{Performances of the models in low and high infectivity scenarios in the BA network}
\label{appendix: BA_Low and high infectivity}

\begin{figure*}[h]
\centering
\includegraphics[scale=0.28]{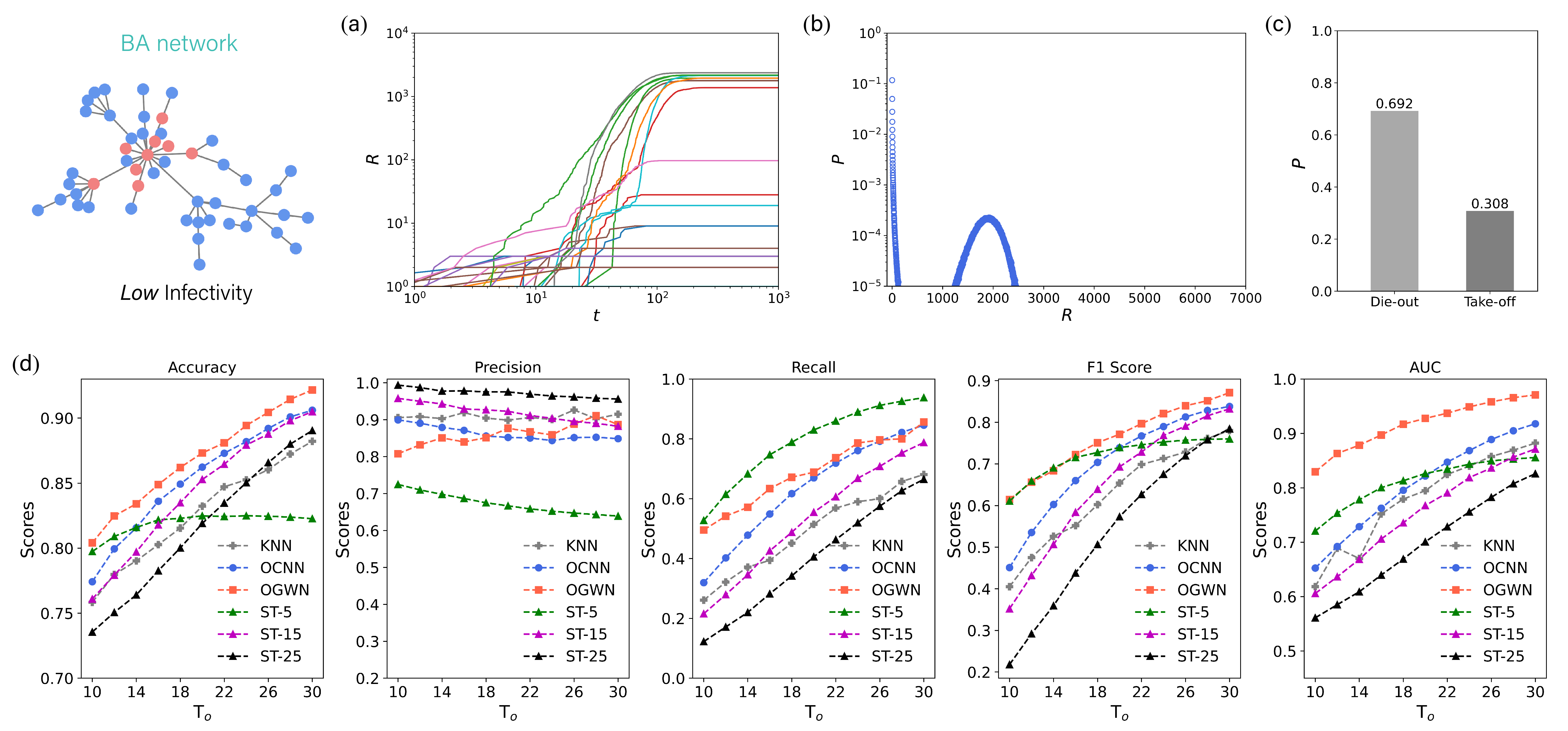}
\caption{Stochastic outbreak prediction with low infectivity scenario. (a) The trajectories of 100 stochastic simulation samples with low spreading infectivity. (b)  The distribution of final recovers of 100000 stochastic simulations. (c) Probability of stochastic die-out and take-off. (d) Model performances across varying observation times  \( T_o \) ranging from 10 to 30.}
\label{fig:BA_low_infectivity}
\end{figure*}

\begin{figure*}[h]
\centering
\includegraphics[scale=0.28]{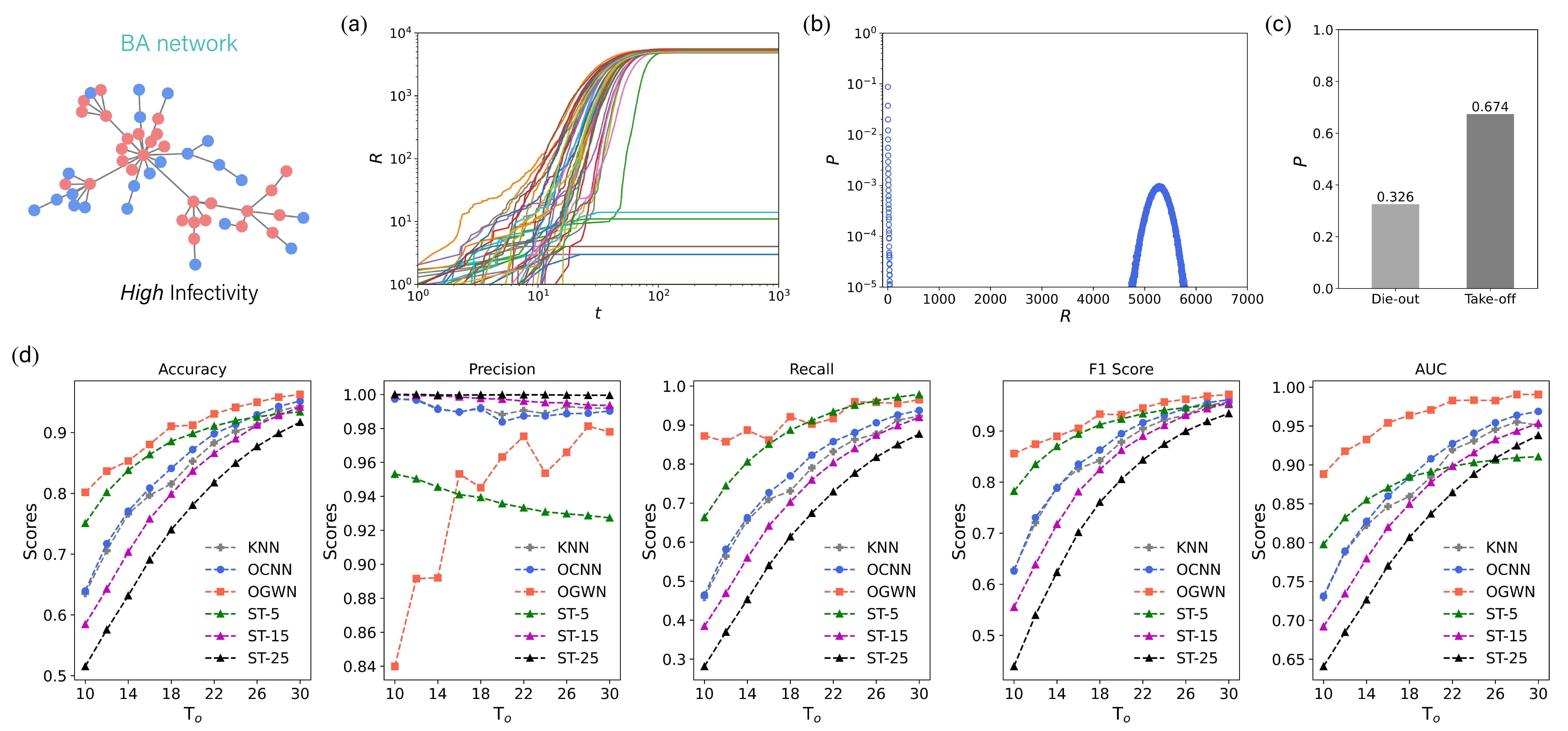}
\caption{Stochastic outbreak prediction with high infectivity scenario. (a) The trajectories of 100 stochastic simulation samples with high spreading infectivity. (b)  The distribution of final recovers of 100000 stochastic simulations with high infectivity. (c) Probability of stochastic die-out and take-off. (d) Model performances across varying observation times  \( T_o \) ranging from 10 to 30.}
\label{fig:BA_high_infectivity}
\end{figure*}

\clearpage
\section{Performances of the models in low and high infectivity scenarios in the ER network}
\label{appendix: ER_Low and high infectivity}

\begin{figure*}[h]
\centering
\includegraphics[scale=0.28]{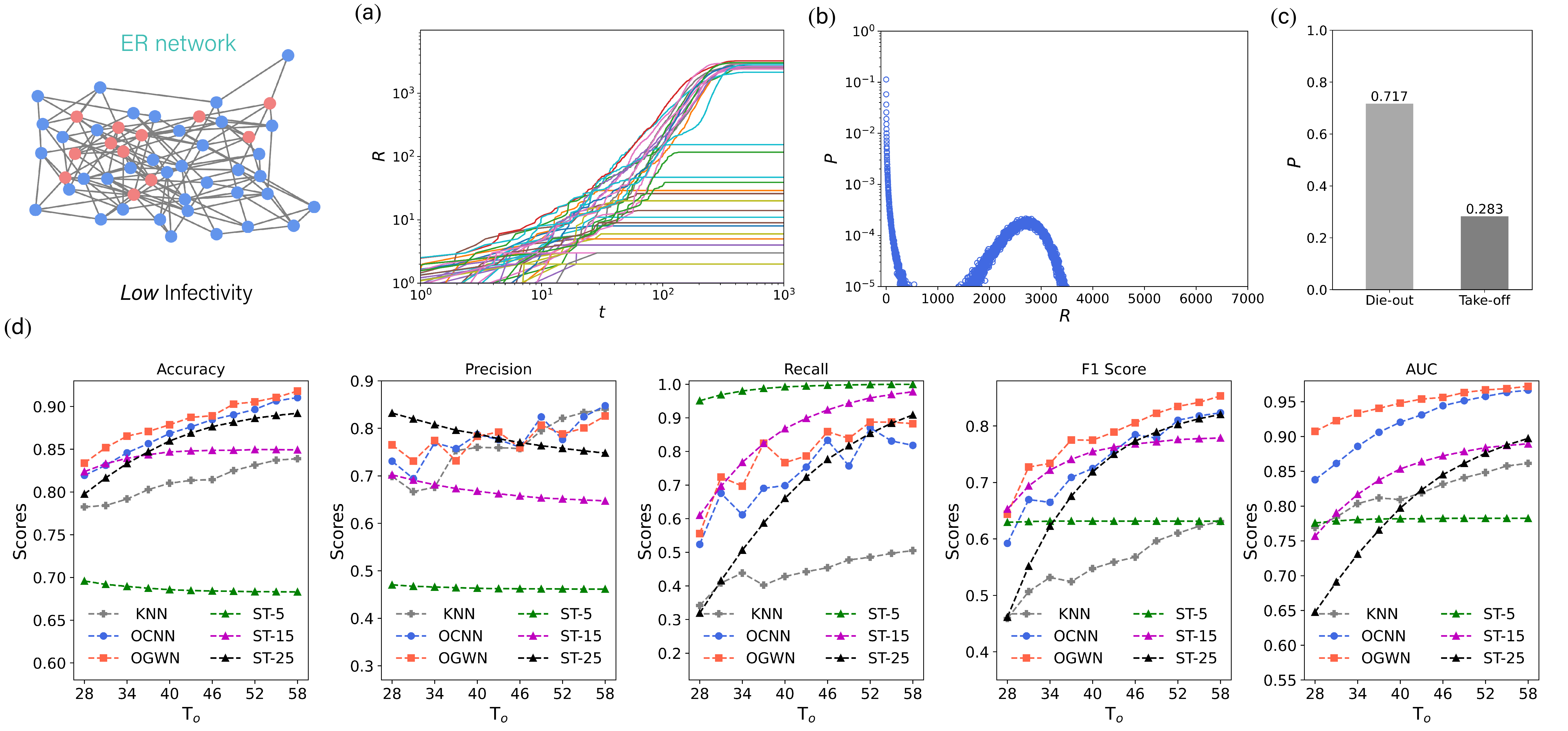}
\caption{Stochastic outbreak prediction with low infectivity scenario in ER network. (a) The trajectories of 100 stochastic simulation samples with low spreading infectivity. (b)  The distribution of final recovers of 100000 stochastic simulations. (c) Probability of stochastic die-out and take-off. (d) Model performances across varying observation times  \( T_o \) ranging from 28 to 58.}
\label{fig:ER_low_infectivity_er}
\end{figure*}

\begin{figure*}[h]
\centering
\includegraphics[scale=0.28]{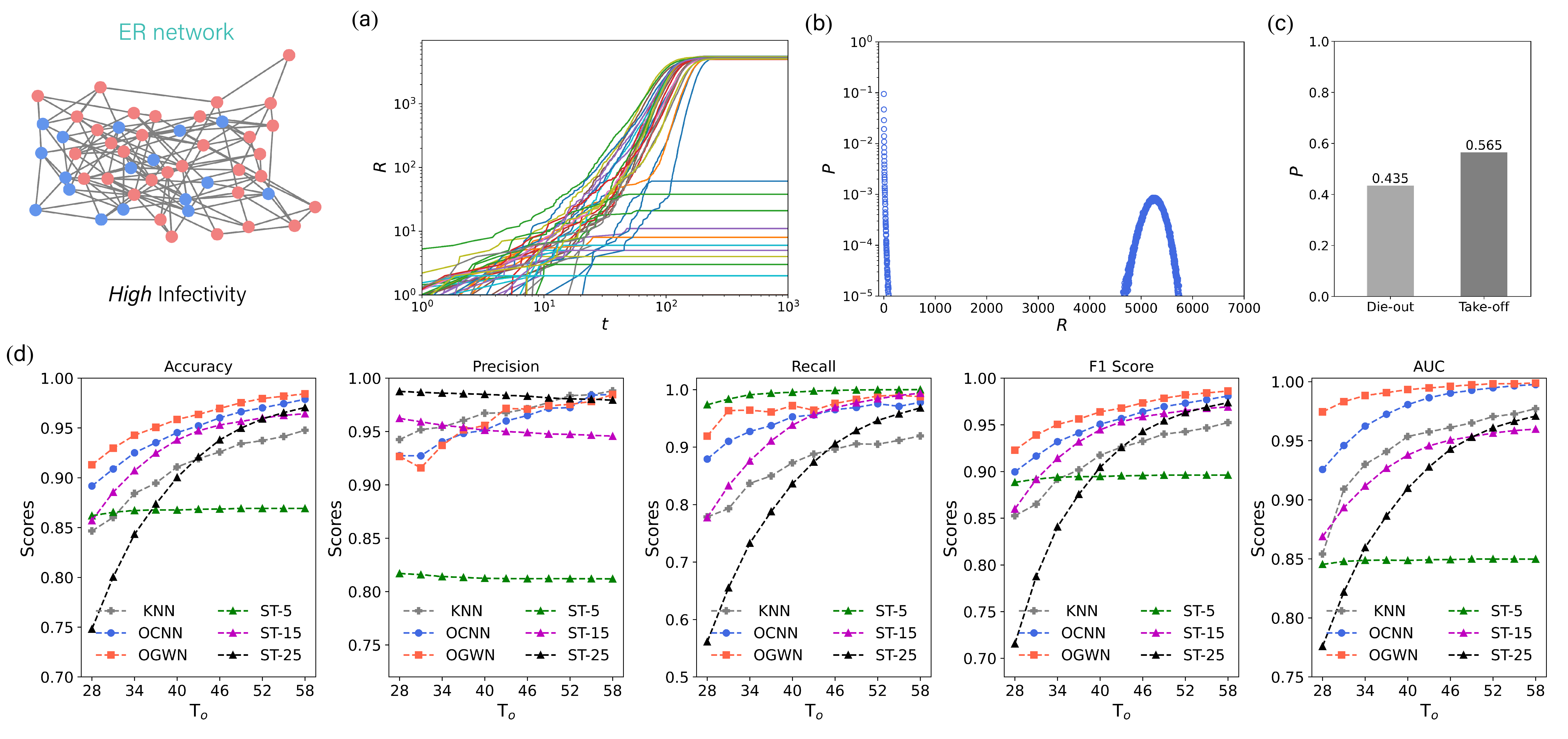}
\caption{Stochastic outbreak prediction with high infectivity scenario in ER network. (a) The trajectories of 100 stochastic simulation samples with high spreading infectivity. (b)  The distribution of final recovers of 100000 stochastic simulations with high infectivity. (c) Probability of stochastic die-out and take-off. (d) Model performances across varying observation times  \( T_o \) ranging from 28 to 58.}
\label{fig:ER_high_infectivity_er}
\end{figure*}

\clearpage

\end{document}